\documentclass[acmtog,screen,
]{acmart}
\usepackage[strings]{underscore}  
\acmSubmissionID{361}
\usepackage{soul}
\usepackage{framed}

\usepackage{listings}
\usepackage{caption}
\usepackage{subcaption}

\definecolor{kc}{rgb}{1,0.9,0.9}

\definecolor{jrk}{rgb}{0.9,0.9,1.0}

\definecolor{tml}{rgb}{0.7,0.9,0.7}

\definecolor{jbt}{rgb}{0.9,0.9,0.9}

\title{Designing Perceptual Puzzles by Differentiating Probabilistic Programs}
\author{Kartik Chandra}
\affiliation{
  \department{Computer Science and Artificial Intelligence Laboratory (CSAIL)}
  \institution{MIT}
  \city{Cambridge}\state{MA}\country{USA}
}
\author{Tzu-Mao Li}
\affiliation{
  \institution{UCSD}
  \city{San Diego}\state{CA}\country{USA}
}
\author{Joshua Tenenbaum}
\affiliation{
  \department{Department of Brain and Cognitive Sciences}
  \department{CSAIL}
  \department{Center for Brains, Minds and Machines}
  \institution{MIT}
  \city{Cambridge}\state{MA}\country{USA}
}
\author{Jonathan Ragan-Kelley}
\affiliation{
  \department{CSAIL}
  \institution{MIT}
  \city{Cambridge}\state{MA}\country{USA}
}



\copyrightyear{2022} 
\acmYear{2022} 
\setcopyright{rightsretained} 
\acmConference[SIGGRAPH '22 Conference Proceedings]{Special Interest Group on Computer Graphics and Interactive Techniques Conference Proceedings}{August 7--11, 2022}{Vancouver, BC, Canada}
\acmBooktitle{Special Interest Group on Computer Graphics and Interactive Techniques Conference Proceedings (SIGGRAPH '22 Conference Proceedings), August 7--11, 2022, Vancouver, BC, Canada}
\acmDOI{10.1145/3528233.3530715}
\acmISBN{978-1-4503-9337-9/22/08}

\begin{document}

\begin{abstract}
We design new visual illusions by finding ``adversarial examples'' for principled models of human perception --- specifically, for probabilistic models, which treat vision as Bayesian inference. To perform this search efficiently, we design a \emph{differentiable} probabilistic programming language, whose API exposes MCMC inference as a first-class differentiable function. We demonstrate our method by automatically creating illusions for three features of human vision: color constancy, size constancy, and face perception.
\end{abstract}


\maketitle

\begin{figure}
    \centering
    \includegraphics[page=4,trim=450 0 450 0,clip,width=\linewidth]{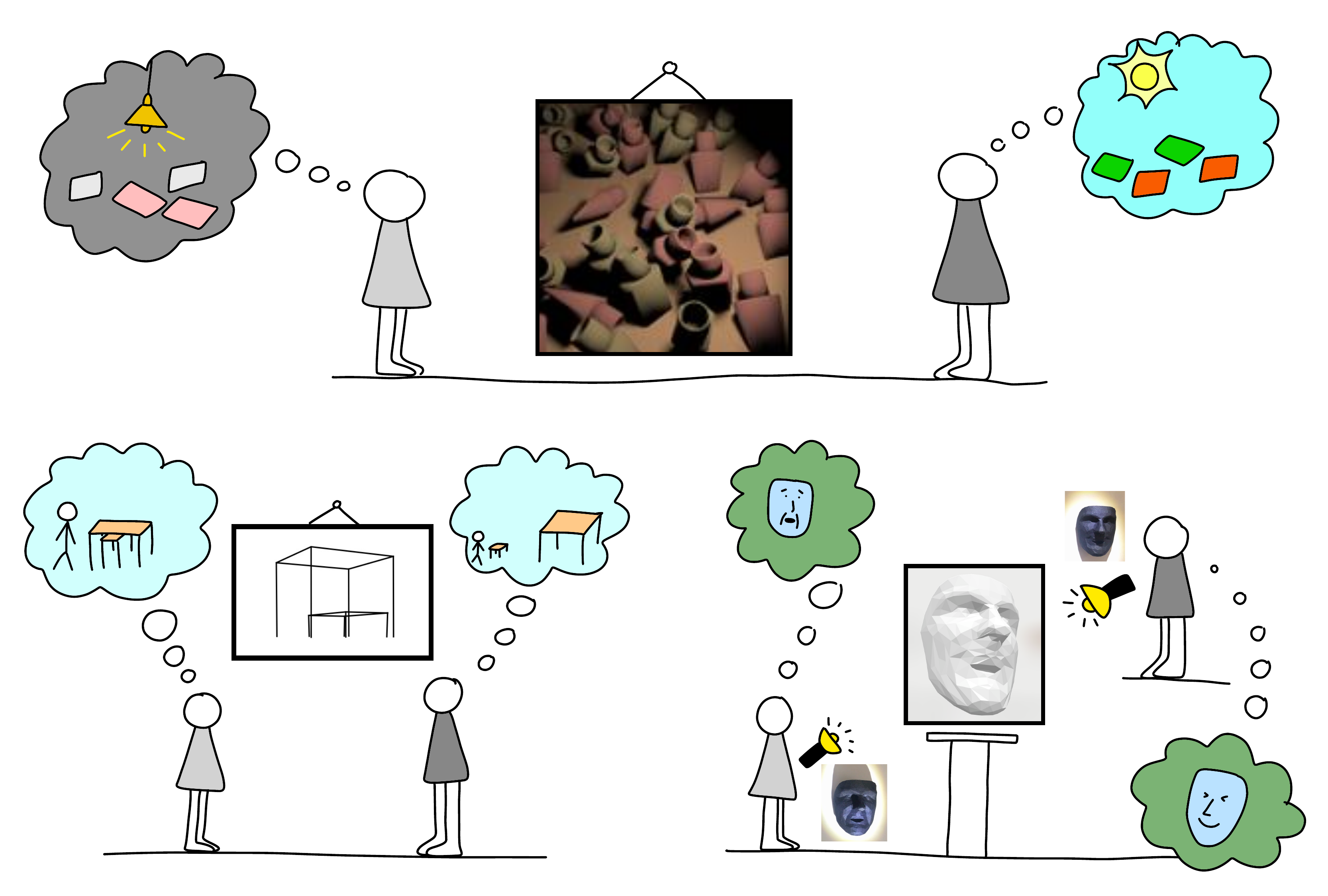}
    \caption{Illusion synthesis: We find scenes where different people perceive different colors (top), where most people infer incorrect geometry (left), and where changing the lighting causes a face to change expression (right). To search for these scenes, we perform gradient descent optimization over Bayesian models of human visual perception, which are implemented in our differentiable probabilistic programming language.}
    \label{fig:teaser}
\end{figure}

\section{Introduction}
\label{sec:introduction}

Human visual perception does not always match objective reality. Consider \href{https://en.wikipedia.org/wiki/The_dress}{``The Dress,''} a visual illusion discovered by chance in 2015. This photograph of a blue and black striped dress elicits a strange perceptual effect: while most viewers indeed describe the dress as blue and black, roughly a third instead confidently describe it as white and golden \cite{lafer2015striking}. Because these two perceptual modes are so dramatically different, viewers are often baffled to learn that others see the same image differently.

Illusions like ``The Dress'' have long been studied for insight into perception. This paper is motivated by the question, \emph{Can we systematically generate new such illusions in a principled manner?} Following \citet{durand2002perceptual}, we would like to think of such perceptually-aware image synthesis as ``inverse inverse rendering.'' Given a model of human visual perception that infers a scene from an image (``inverse rendering''), we wish to search for input images that elicit interesting responses as output from the model (``\emph{inverse} inverse rendering'').

What model should we choose?
A long line of cognitive science research suggests that human visual perception is modeled well by \emph{probabilistic models}, which treat perception as Bayesian inference. Under such a model, a viewer has some \emph{prior belief} about the statistical distribution of scenes (objects, their colors, and lighting conditions) in the world. Faced with an \emph{observation} (an image), the viewer performs \emph{inference} to update their prior belief into a \emph{posterior belief} using Bayes' rule:
\(
p(\text{Scene} \mid \text{Image}) \propto {p(\text{Image} \mid \text{Scene})p(\text{Scene})}
\).

Probabilistic models have already been used for inverse rendering in a variety of settings. Here, however, we are interested in \emph{inverse} inverse rendering: we seek an $\text{Image}^\star$ such that the model's inferred posterior belief, given by the distribution $p(\text{Scene} \mid \text{Image}^\star)$, is confidently incorrect. In this way, $\text{Image}^\star$ is like an ``adversarial example'' --- not for a deep neural network, but instead for a small, principled model of one aspect of human perception.

In this paper, we offer a general-purpose tool for solving such inverse problems: a \emph{differentiable probabilistic programming language}. Regular probabilistic programming languages (PPLs) allow users to express probabilistic models as structured generative processes, which are then inverted into efficient programs for performing Bayesian inference with respect to a given observation. By enabling differentiation through the inference process, a \emph{differentiable} PPL allows users to optimize (by gradient descent) an observation to evoke some desired property in the inferred posterior distribution. Using this tool, we can model a variety of perceptual phenomena with simple generative models --- just a few lines of code --- and then search for stimuli that elicit interesting behaviors from the models.

This work is a synthesis of many ideas from a variety of fields. Section~\ref{sec:related-work} reviews necessary background, situating our work in the broader context of current machine learning and cognitive science research. After that, we present our two contributions: a differentiable PPL (Section~\ref{sec:implementation}), and a formalization of ``illusion synthesis'' as an application of differentiable PPLs (Section~\ref{sec:applications}). We apply our method to create illusions in three domains: color constancy~(\ref{sec:color-constancy}), size constancy~(\ref{sec:size-constancy}), and face perception~(\ref{sec:face-constancy}).
Finally, we reflect on our work's broader implications for the graphics and cognitive science communities, and discuss scope for future work (Section~\ref{sec:future-work}).
\section{Background}
\label{sec:related-work}

\subsection{Motivation: Why do we need Bayesian models?}
Because our goal is to find visual illusions that work for humans, we need a model of vision whose edge-cases match those of humans. This choice is not straightforward: convolutional neural networks (CNNs), for example, do not meet this criterion because even though we can easily fool them with adversarial examples \cite{szegedy2013intriguing}, nearly a decade of research teaches us that those examples do not transfer to fool humans \cite{sinz2019engineering}. Indeed, this is precisely why adversarial examples are intriguing: they demonstrate how deep learning differs fundamentally from human perception.

While this consensus is occasionally disputed, the evidence still does not suggest any method for generating robust, compelling illusions like ``The Dress.'' For example, \citet{elsayed2018adversarial} optimized images to be misclassified by an ensemble of ten CNNs, and found that the resulting adversarial examples could also cause humans to select the same incorrect labels. However, the humans were only fooled in the ``time-limited'' setting where images were flashed for just 60-70ms --- the effect disappeared for longer presentations. Similarly, \citet{zhou2019humans} found that humans could pick out the incorrect label that a fooled CNN would assign. However, this was only in a forced-choice setting where participants were given a constrained set of incorrect labels to choose from. Neither of these effects is like ``The Dress,'' which persists even under natural viewing and reporting conditions.

Seeking to capture exactly these robust perceptual effects, we turn away from CNNs, and instead consider an alternate model of vision: the \emph{Bayesian} model, which seeks to explain human vision as the computations of a rational perceptive agent. Below, we provide a brief introduction to Bayesian models, why we expect them to match human perception, and how they are implemented in practice (we direct readers seeking further background to a textbook by \citet{goodman2016probmods2}). Finally, we discuss algorithms for differentiating over those models, which in turn enable us to optimize ``adversarial examples'' that indeed transfer to humans.

\subsection{Modeling vision as Bayesian inference}
\label{sec:vision-as-bayesian-inference}

\paragraph{Probabilistic models of perception} A long line of work from the cognitive science community (see, e.g., a book by \citet{knill1996perception} or surveys by \citet{mamassian2002bayesian} and \citet{kersten2004object}) has argued that because many possible scenes can map to the same image on the human retina, parsing an image into a scene by ``inverse graphics'' is an ill-posed problem. Therefore, vision must be probabilistic in nature: it must rationally infer probable scenes from an image based on prior beliefs about the statistics of scenes in the viewer's environment. For example, this view motivates a simple Bayesian model of color constancy \cite{brainard1997bayesian}: We start with a distribution $p(\text{Light}, \text{Color})$ representing our prior belief over scenes (for example, that daylight is likelier than bright green light), and a distribution $p(\text{Image} \mid \text{Light}, \text{Color})$ that encodes how scenes are rendered into images along with any uncertainty in that process (for example, sensor noise in the eye or camera). Bayes' rule then allows us to infer the posterior distribution of scenes, conditioned on a given observed image: $p(\text{Light}, \text{Color} \mid \text{Image}) \propto p(\text{Light}, \text{Color})p(\text{Image} \mid \text{Light}, \text{Color})$.

This account of vision posits a natural ``specification'' of what vision computes, independent of what the algorithmic implementation or biological realization of this specification may be \cite{marr1982vision}. A variety of recent systems have sought to apply this insight to build artificial models of visual scene perception \cite{kulkarni2015picture, mansinghka2013approximate, gothoskar20213dp3}. Ahead of time, the authors of these systems define a probability distribution over scenes in their target domain, expressing their prior belief about the statistics of natural scenes. When given an input image, the systems condition their prior distribution on having observed that image. They then apply Bayes' rule to infer the posterior distribution of scenes that best explain the image. Such models can account for a variety of perceptual phenomena in humans, including some illusions \cite{weiss2002motion, geisler2002illusions}.

\paragraph{Probabilistic programming languages (PPLs)} PPLs are domain-specific languages that enable users to express complicated probability distributions by implementing generative processes that sample random variables as they execute \cite{van2018introduction, carpenter2017stan}. For example, we could describe the distribution $p(\text{Image} \mid \text{Light}, \text{Color})$ in a PPL by writing a small graphics engine that randomly samples camera parameters, and then uses those parameters to render an image of the scene using the given light and color. The PPL compiles this generative procedure into a function $q$ that evaluates the joint probability density of a given concrete instantiation of the random variables. PPLs also allow users to express conditional distributions by fixing the values of the observed random variables.

\paragraph{Performing inference using MCMC sampling} Once we have a way to evaluate the conditional probability density, we would like to get a concrete sense of the distribution --- for example, approximate its mean and variance --- by drawing samples from it. A standard technique for sampling from a probability distribution is the Markov Chain Monte Carlo (MCMC) algorithm: We perform a random walk over the domain of $q$, choosing transition probabilities such that the random walk's stationary distribution is proportional to $q$. For distributions with continuous domains, a common choice of transition probabilities is given by the Hamiltonian Monte Carlo (HMC) algorithm~\cite{duane1987hybrid, neal2011mcmc}. HMC treats the probability density $q$ as a physical potential energy well and simulates a randomly-initialized particle's trajectory in that well for a fixed time window. Intuitively, such a particle would indeed spend longer in regions of low potential (or high probability), exactly as desired.

\subsection{Differentiating inference}
\label{sec:differentiating-inference}

Now that we can approximate expectations taken over conditional probability distributions, we would like to differentiate these expectations with respect to parameters of the distribution. Fortunately, because each step of the HMC random walk is given by a physics simulation with continuous dynamics, we can simply apply automatic differentiation to compute derivatives of \emph{samples} with respect to parameters of the probability distribution.\footnote{HMC has one non-differentiable component: the ``accept/reject'' step or ``Metropolis-Hastings correction,'' which accounts for numerical imprecision in the physics simulation. Our work, along with the others referenced in this section, elides this step to make the sampler fully differentiable, even though it comes at the cost of slightly biasing the samples.}

In the language of statistical machine learning, this technique is analogous to the ``reparametrization trick''  \cite{kingma2014autoencoding} or ``pathwise gradients'' technique \cite{mohamed2020monte} where the ``reparametrization'' or ``path'' maps HMC's random inputs (i.e. the particle's random initializations) to a sample from the target distribution. Indeed, to the best of our knowledge, this technique was first proposed by the ML community, in the context of variational inference \cite{salimans2015markov}. More recently, it has been used to train energy-based models \cite{du2021improved, dai2019exponential, vahdat2020undirected, zoltowski2020slice}, for Bayesian learning \cite{zhang2021differentiable}, and to optimize the hyperparameters of HMC itself \cite{campbell2021gradient}.

In the above settings, the target probability distribution is typically parametrized by a black-box model (e.g. a deep neural network) whose weights are to be optimized. Our work, in contrast, applies these ideas to differentiate through richly-structured generative models expressed in PPLs (with embedded renderers/simulations), focusing specifically on perceptual models.


\subsection{Space-efficient differentiation via reversible dynamics}
\label{sec:reversible-dynamics}

The algorithm as described above does not scale to our setting because it uses too much memory. Backpropagation requires storing all the intermediate values computed by the target function, and HMC sampling produces a tremendous amount of intermediate data because it has two nested loops (ranging over samples and steps of the physics simulation). For our applications, which involve rendering within the generative model, HMC's total memory usage quickly multiplies beyond what is practical.
Our solution is to apply ``reversible learning,'' a technique originally developed for hyperparameter optimization in deep learning \cite{maclaurin2015gradient}, also later used in variational inference \cite{zhang2021differentiable}. Rather than \emph{storing} intermediate values, we \emph{recompute} them dynamically during backpropagation by running the physical simulation in reverse, backwards through time.
\section{Differentiable probabilistic programming}
\label{sec:implementation}

Section~\ref{sec:related-work} reviewed the cognitive-science and algorithmic foundations of our work. Now, we will briefly put aside questions of perception, and provide a tour of our key tool: the differentiable probabilistic programming language, which augments regular PPLs with the ability to differentiate through inference. To keep our discussion concrete, we will adopt the following small example problem:
\begin{oframed}
\noindent
    In summer, the daily high temperature $T$ in San Francisco is distributed normally, with mean $70^\circ$ and standard deviation $5^\circ$. Your old thermometer reports a measurement $M$ of the true temperature, with some added Gaussian noise of $\sigma=2^\circ$.
    \begin{enumerate}
        \item On a hot day, your thermometer reports $100^\circ$. Knowing your prior belief that $T \sim N(70, 5)$ and $M \sim N(T, 2)$, what do you think the true temperature is?
        \item What measurement would the thermometer have to report for you to believe the temperature is truly $100^\circ$?
    \end{enumerate}
\end{oframed}
Question (1) is a standard Bayesian inference problem that seeks to invert the observed measurement into an inferred temperature; it is easily solved by existing PPLs. Question (2) is the ``inverse inverse problem'' that corresponds to (1) and requires our differentiable PPL. (In the same way, we will later treat illusion synthesis as an ``inverse inverse problem'' corresponding to inverse graphics.)

In this simple Gaussian setting, we can analytically solve both problems in closed form. For Question (1), we have $E[T \mid M = 100^\circ]=(2780/29)^\circ \approx 95.86^\circ$, and for question (2), the solution to the equation $E[T \mid M=m] = 100^\circ$ is $m=524/5^\circ = 104.8^\circ$. Below, we will use our differentiable PPL to approximate these values numerically.

\subsection{Writing a generative model}

The first step is to encode our problem setup into a probabilistic generative model. Following languages like Anglican \cite{wood2014new}, we provide two important primitives in our PPL: \lstinline{sample}, which yields a fresh, independent sample from a given distribution, and \lstinline{observe}, which conditions the generative process on a sample from a distribution being equal to a given observed value.
\begin{lstlisting}
def model(M):
  sample T ~ N(70, 5)
  observe M from N(T, 2)
  return T
\end{lstlisting}
Here, we define a model that can be conditioned on $M$. The model samples a true temperature $T$ from the given distribution, then observes that the noisy measurement is equal to $M$. Finally, it returns $T$ to signal that it is the value to be inferred.

This generative process implicitly defines a joint probability distribution over the random variables in the model. In particular, applying Bayes' rule, we have
\(
p(t, m) = p(t)p(m \mid t) = \Phi\left(\frac{t - 70}{5}\right)
\Phi\left(\frac{m - t}{2}\right)
\)
where $\Phi$ is the Gaussian probability density function. More generally, each \lstinline{sample} and \lstinline{observe} statement appends a multiplicative factor to the joint density. The PPL automatically compiles the generative process into a function that evaluates the (unnormalized) conditional density, given values for the latent random variables and the observations on which to condition.

\subsection{Performing MCMC inference using HMC}

Our language provides a function \lstinline{hmc_sample}, which takes as input the HMC hyperparameters ($N$, the number of samples to take; $L$, the number of iterations of the physical simulation; and $\epsilon$, the epsilon value used by the simulation's integrator) and the observation on which to condition. The appropriate HMC hyperparameters vary by model, and selecting them typically requires some trial and error.

The function \lstinline{hmc_sample} performs HMC inference and returns an array of $N$ samples from the conditional distribution (of the random variable $T$, in our model). Typically, it is helpful to skip the first few samples, which are biased by the sampler's ``burn-in period,'' during which the initial conditions' influence has not yet been washed out by the random walk. We \lstinline{skip} the first 100 samples for this reason.
\begin{lstlisting}
samples = hmc_sample(
  N=1000, L=300, eps=0.01, skip=100,
  M=100  # observation on which to condition
)
print(samples.mean())
\end{lstlisting}
This outputs the approximation $E[T \mid M=100] \approx 95.959^\circ$ (the analytical solution is $95.86^\circ$).

For brevity, from here on we will elide the HMC hyperparameters and the \lstinline{.mean()} by defining the convenience function \lstinline{infer}, which simply calls \lstinline{hmc_sample} with appropriate hyperparameters and takes the mean of the resulting samples. Thus, the above lines of code can be shortened to \lstinline{print(infer(M=100))}.

\subsection{Solving the ``inverse inverse problem'' by gradient descent}

Finally, we are ready to solve the ``inverse inverse problem'': finding an $m$ such that the inferred value of $T$ is exactly $100^\circ$. To compute this value by gradient descent optimization, we must write down an optimization target. Here, we seek to minimize the difference between the inferred $E[T \mid M = m]$ and the desired value of $100^\circ$.
\begin{lstlisting}
def loss(m):
  return (infer(M=m) - 100) ** 2
\end{lstlisting}
Because \lstinline{infer} is end-to-end differentiable (see Sections~\ref{sec:differentiating-inference} and \ref{sec:reversible-dynamics}), we can simply apply automatic differentiation and backpropagate through the computation of \lstinline{loss}. After 100 steps of gradient descent with a step size of 0.1, the optimization converges, and we obtain our solution $m = 104.80379^\circ$ (the analytical solution is $104.8^\circ$).
\begin{figure*}
     \centering
     \includegraphics[width=\linewidth]{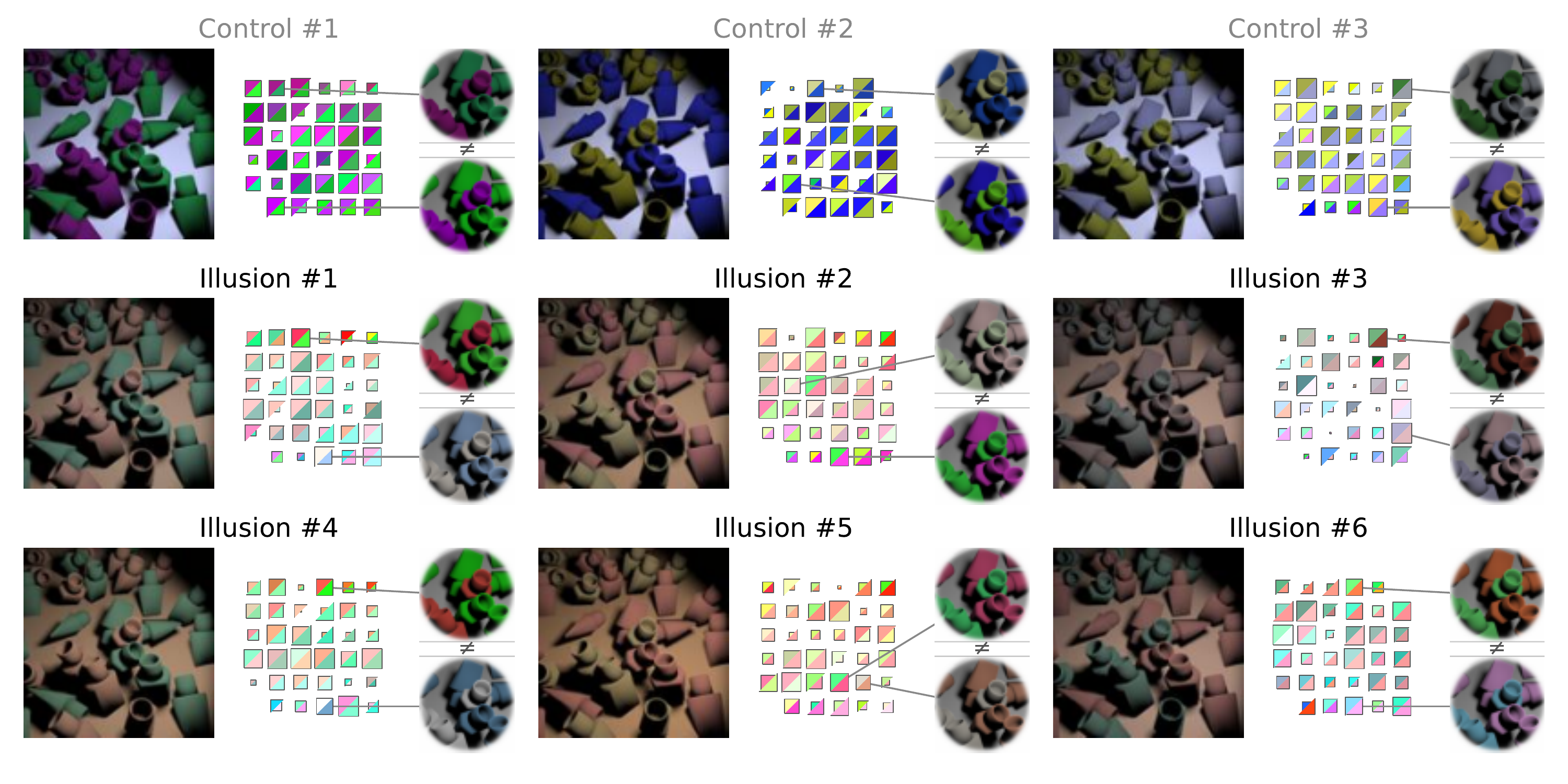}
    \caption{Color constancy illusions created by our method (Section~\ref{sec:color-constancy}). Each plot shows the image we presented to participants in our study, along with swatches representing the colors that they reported for the two kinds of erasers. The size of each half-square is proportional to the participant's self-reported confidence in that color response. Selected high-confidence responses ($\geq 7/10$ for both colors) are shown in circled insets, re-lit in neutral light as the participants themselves saw in the guide display when picking colors. For controls, we selected the most distant pairs in CIELAB space; for illusions, we handpicked responses to represent significant perceptual modes.
    \textbf{The reported colors for the illusions vary dramatically in hue, while the reported colors for the controls (top row) are tightly clustered among perceptually similar hues, only varying slightly in saturation.}
    A small fraction of participants swapped the two colors in their responses. This does not affect the experiment's results, and we present the raw data without manually correcting their errors.
    \emph{Note: This figure is best viewed on a bright, high-resolution color display.}
    }
    \label{fig:color}
\end{figure*}

\section{Applications}
\label{sec:applications}

We are now ready to apply differentiable PPLs to find adversarial examples for probabilistic models of perception by gradient descent. The computations described below were implemented in Python using the JAX automatic differentiation library and run on an NVIDIA TITAN X GPU. Each result is reproducible in minutes using the source code included in the supplementary materials and online at \url{https://people.csail.mit.edu/kach/dpp-dpp/}.

\subsection{Color constancy} 
\label{sec:color-constancy}

We begin with our motivating example of synthesizing new color constancy illusions like ``The Dress.'' To do so by ``inverse inverse rendering,'' we need a probabilistic model and an optimization target.

Our probabilistic model samples scenes that contain colored objects and a single light source. For the light source, we randomly sample a color temperature from $N(6500K, 1000K)$ and uniformly select its brightness from a large range. For the objects, we choose to render colored wedge-top pencil erasers, for three reasons: First, erasers are highly Lambertian, which minimizes lighting cues given away by specular highlights. Second, the distinctive shape of wedge-top erasers makes it easy to recognize the objects as erasers. Finally, erasers come in a wide variety of colors, which allows us to use a simple uniform prior over object color. We fix all other scene parameters, such as the positions of the camera and objects.

To condition our model on an input, we follow \citet{kulkarni2015picture} and add a small amount of pointwise Gaussian noise at each pixel of the rendered image, then assert that the result is the observed image. In our PPL, all of this is expressed as follows:
\begin{lstlisting}
def model(observed_img):
  sample temp ~ N(6500, 1000)
  sample brightness ~ U(0, 2)
  light = planck_to_RGB(temp) * brightness
  sample color1[r,g,b], color2[r,g,b] ~ U(0, 1)
  
  img = render(light, color1, color2)
  observe img from N(observed_img, 0.1)
  return color1, color2
\end{lstlisting}

Finally, for our optimization target, we follow the usual ``adversarial examples'' recipe: we optimize for an image that causes the model to infer colors as different from the true colors as possible.\footnote{Our implementation adds an additional term to the loss that encourages \texttt{color1} and \texttt{color2} to be different, which yields more visually interesting results.}
\begin{lstlisting}
def loss(light, true_color1, true_color2):
  perceived_color1, perceived_color2 =
    infer(render(light, true_color1, true_color2))
  return
    -length(true_color1 - perceived_color1) +
    -length(true_color2 - perceived_color2)
\end{lstlisting}
Why should this optimization target yield \emph{diverse} percepts, as opposed to simply consistently-wrong ones? Recall that human viewers vary in their perceptual priors, as evidenced by ``The Dress'' \cite{wallisch2017illumination}. However, our model only has a single, fixed-but-reasonable prior. We thus expect viewers whose priors match our model to be ``fooled'' by the adversarial examples, and viewers with different priors to perceive the colors correctly. Indeed, when we evaluated our illusions with human viewers, we observed diverse responses rather than consistent ones, as discussed below.

\paragraph{Evaluation}
Because phenomena like ``The Dress'' depend on diversity in perceptual priors, they are not apparent to individuals in isolation: they are only revealed when a group of people disagree about their percepts. Thus, to evaluate our illusions, we conducted an online human subject study.\footnote{This study was conducted with IRB approval, and all subjects were compensated for their time.}

First, we performed the optimization described above for 30 random seeds. For most seeds, gradient descent converged to one of three ``classes'' of illusions. We hand-picked two representative samples from each class to evaluate, for a total of 6 illusions. Next, we recruited 60 na\"ive participants and presented each one with our 6 illusions in randomized order. Randomly-selected controls (images with poor loss in our model) were interspersed every 3 images. Participants reported the colors of the two kinds of erasers using color pickers. They also reported their confidence using sliders labeled ``just guessing'' to ``very confident.'' A guide display re-rendered their selections under neutral lighting in real time. Participants were instructed to disable blue-light filters and the study was conducted during daytime hours in the continental US. Only one participant self-reported any color vision deficiency. We discarded responses from participants who failed attention checks (e.g. ``what did the sliders measure?''), leaving 35 reliable responses, shown in Figure~\ref{fig:color}.
Among these responses, we found that the colors reported for the controls clustered tightly among perceptually similar colors, while the colors reported for the illusions indeed varied widely, as desired.

\paragraph{Other approaches} We end this section with a brief discussion of related work on synthesizing Dress-like illusions. \citet{witzel2020make} transplant the original colors of ``The Dress'' onto new objects and show that this preserves the illusion. However, their method cannot find new ``color schemes'' as ours does. \citet{wallisch2019disagreeing} photograph colorful slippers and white socks under colored light that makes the slippers look gray. Most viewers correctly perceive colored slippers with white socks; however, some instead perceive gray slippers with colored socks. This illusion was cleverly handcrafted with respect to specific qualitative priors (i.e. socks are typically white), and required careful manual tuning to perfect. In comparison, our method automatically and efficiently optimizes illusions under principled natural priors. Finally, our prior work explores a similar effect in \emph{speech} perception, the ``Laurel/Yanny'' illusion, by using heuristic search and large-scale crowdsourcing to find new audio clips that exhibit the same phenomenon \cite{chandra2021beyond}. Our present work seeks to directly synthesize illusions, rather than finding them among natural examples.
\subsection{Size constancy}
\label{sec:size-constancy}

\begin{figure*}
    \centering
    \begin{subfigure}[h]{\textwidth}
    \includegraphics[page=2,trim=0 440 0 0,clip,width=\linewidth]{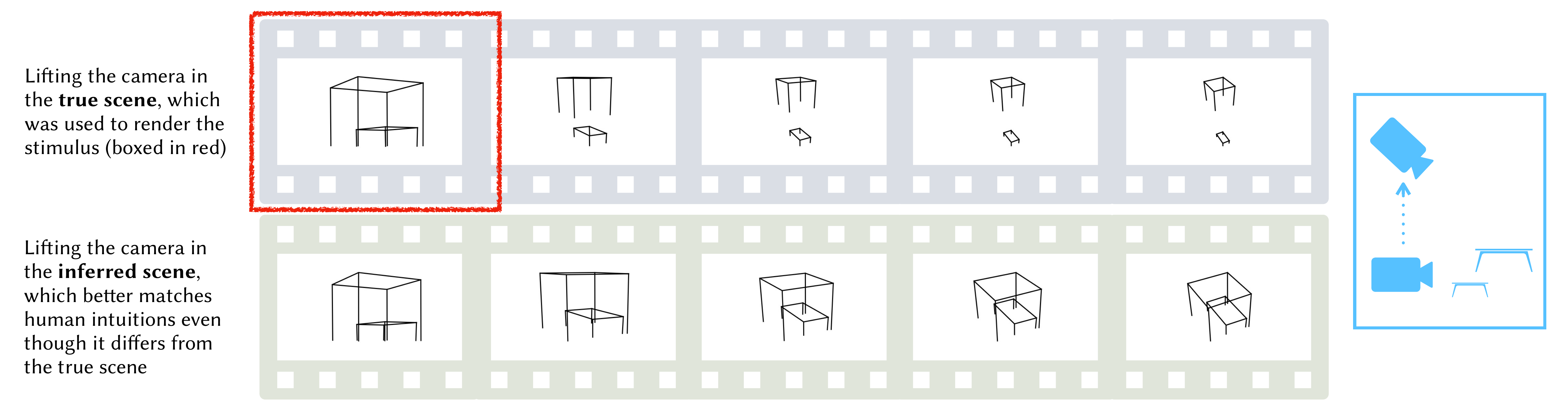}
    \caption{Creating size constancy illusions (Section~\ref{sec:size-constancy}). The synthesized stimulus, outlined in red, is an image of two far-apart tables, as revealed by the camera motion in the upper (blue) film. However, due to our prior belief about the statistical distribution of real-world table sizes, we intuitively perceive the image to show two stacked tables instead: this is shown in the lower (green) film, which applies the same camera motion to the scene inferred by our model.}
    \label{fig:size-constancy-1}
    \end{subfigure}
    \begin{subfigure}[h]{0.35\textwidth}
    \centering
    \includegraphics[page=3,trim=0 0 1600 0,clip,width=\linewidth]{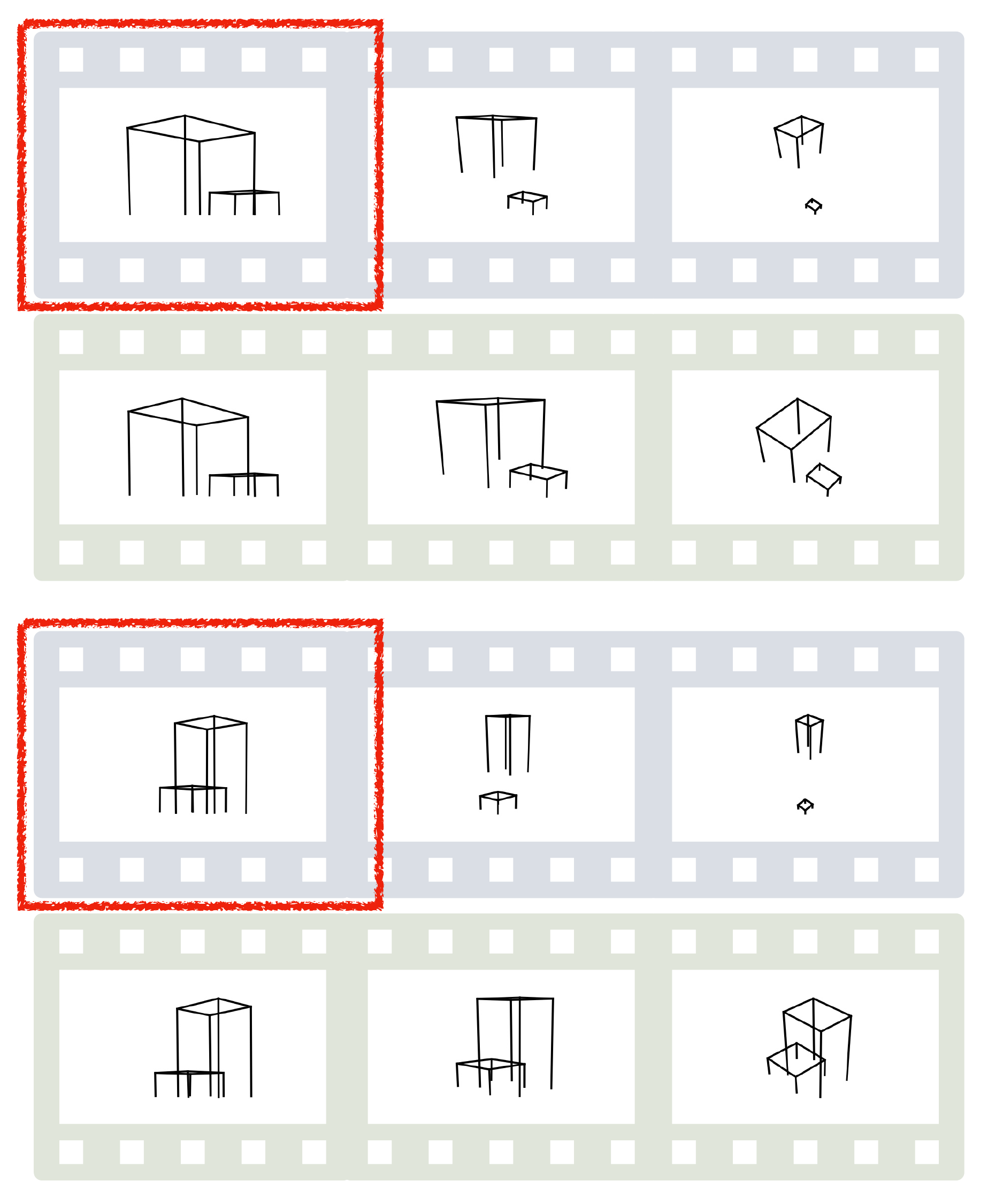}
    \caption{Two more automatically-generated examples of the size constancy illusion described in Section~\ref{sec:size-constancy}.}
    \label{fig:size-constancy-2}
    \end{subfigure}
    \hfill
    \begin{subfigure}[h]{0.6\textwidth}
    \centering
    \includegraphics[page=3,trim=900 0 0 0,clip,width=\linewidth]{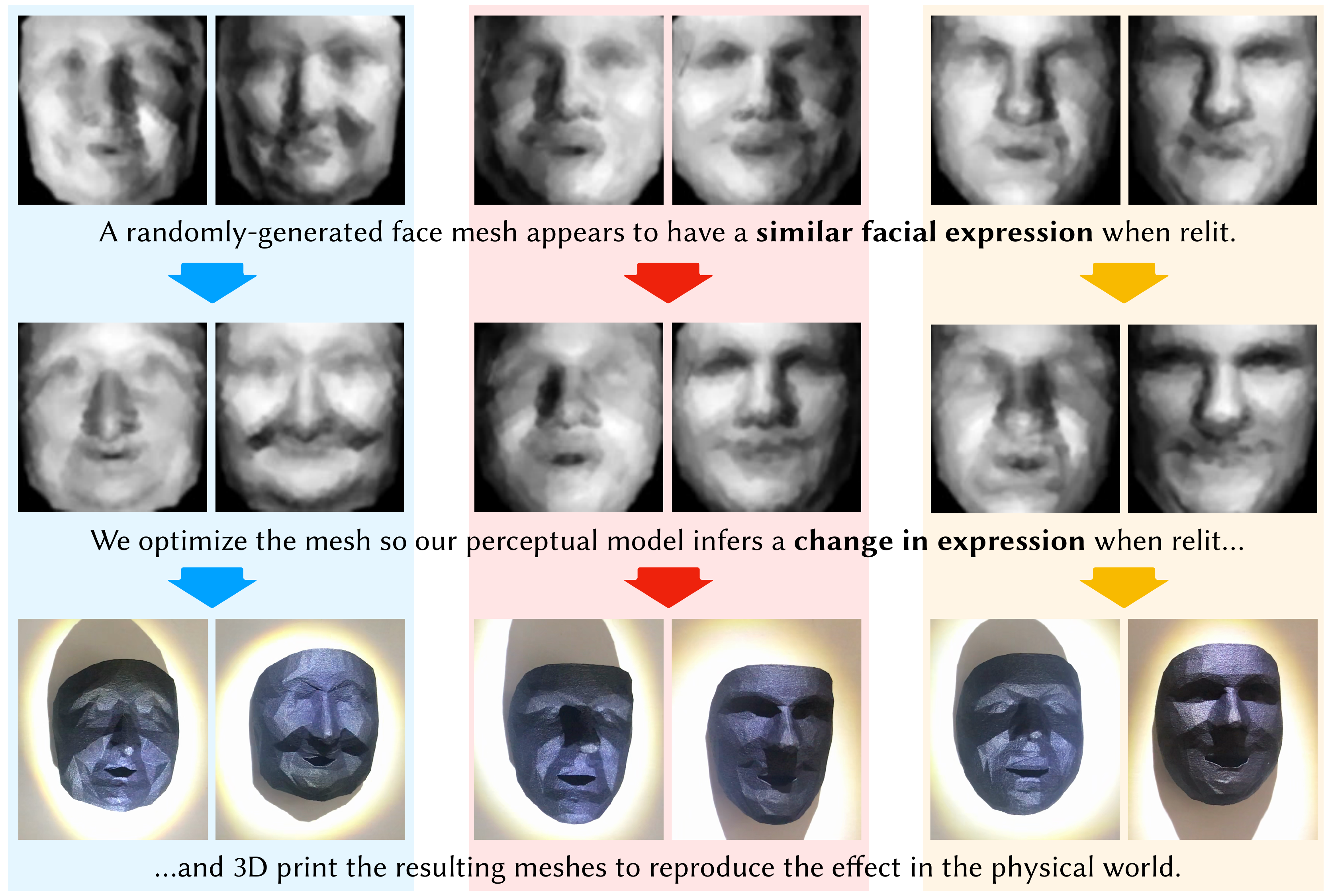}
    \caption{Creating face perception illusions (Section~\ref{sec:face-constancy}). Each pair of images shows the same mesh lit from two different directions.}
    \label{fig:face}
    \end{subfigure}
    \caption{More illusions created by our method.}
\end{figure*}

Another ``constancy'' in human visual perception is size constancy: we infer objects to be the same size, even when they are viewed from different perspectives. Famous illusions of size constancy include the Ponzo illusion and the ``Shepard Tables'' \cite{shepard1994perceptual}.
Here, we try to produce illusions of \emph{forced perspective}, where an unusual camera angle is used to make an object appear larger or smaller (or farther or nearer) than it truly is. This technique is often used by architects to enhance a building's grandeur; oppositely, it is often used by tourists to take souvenir photographs of themselves appearing to hold monuments in their hands.

First, we create a generative model for scenes containing a set of reasonably-sized tables, observed by a randomly-initialized camera. As in previous examples, the model is then conditioned on an observed image. (In this model, we express ``images'' as ordered sequences of vertices representing a 2D projection of the tables' corners.) We seek to infer each table's position and size.
\begin{lstlisting}[mathescape]
def model(observed_tables[]):
  # Camera position (cylindrical coordinates)
  sample r ~ U(8,16); $\theta$ ~ U(0,$2\pi$); h ~ N(0,3)
  camera = make_camera_at(r, $\theta$, h).look_at(origin)

  for i, obs in observed_tables:
    sample size[x,y,z] ~ U(.5,4); pos[x,y] ~ N(0,2)
    tables[i] = make_table(size, pos)
    image = perspective_project(camera, tables[i])
    observe obs from N(image, 0.01)
  return tables
\end{lstlisting}

Next, we express our optimization target. We are looking for an arrangement of tables and camera parameters such that the first table \emph{appears} nearer than it truly is: that is, such that the position inferred by our model is closer to the origin than the true position with which the image was generated.
\begin{lstlisting}
def loss(camera, true_tables[]):
  observed_tables =
    perspective_project(camera, true_tables)
  inferred_tables = infer(observed_tables)

  return
    length(inferred_tables[0].position) -
    length(true_tables[0].position)
\end{lstlisting}
Finally, we optimize this loss by gradient descent.
Some examples of optimized illusions are shown in Figures~\ref{fig:size-constancy-1} and \ref{fig:size-constancy-2}. Both our model, and human intuitions, are ``fooled'' by these images.
\subsection{Face perception}
\label{sec:face-constancy}

For our last application, we were inspired by two effects related to human face perception. First, \citet{troje1998illumination} observe that merely changing the direction of lighting on a human face can change a viewer's perception of the face's orientation. Second, \citet{oliva2006hybrid} show that small changes to a carefully-constructed ``hybrid image'' can change a viewer's perception of the face's expression, an effect previously observed by \citet{livingstone2000warm} in relation to the Mona Lisa. Hoping to combine these two effects to design a new illusion, we asked: \emph{Can we find faces that appear to change their \emph{expressions} in different lighting conditions?}

Following \citet{kulkarni2015picture}, we built a simple probabilistic model of face perception using the Basel Face Model \cite{gerig2018morphable}, a 3D morphable model that outputs meshes of human faces given low-dimensional latent vectors encoding identity and expression. Our generative model samples latent vectors for the morphable model independently and shines a randomly-oriented directional light on the output mesh. This resulting scene is then rendered using the SoftRas differentiable rendering algorithm~\cite{liu2019soft}. As in Section~\ref{sec:color-constancy}, we observe an input image by adding pointwise Gaussian noise to each pixel.
\begin{lstlisting}
def model(observed_img):
  sample identity[0:n], expression[0:n] ~ N(0, 1)
  mesh = BaselFaceModel(identity, expression)
  sample lighting_dir[x,y,z] ~ N(0, 1)
  img = SoftRas(mesh, lighting_dir)
  observe img from N(observed_img, 0.1)
  return expression
\end{lstlisting}
Some examples of unconditional samples from this model are shown in the top row of Figure~\ref{fig:face}. So far, the facial expressions appear mostly the same even if the lighting direction is changed. Indeed, our model is easily able to infer the expressions in all of these images.

Next, we wrote an optimization target to search for a face mesh and lighting direction such that the face ``looks'' different to our model if the lighting direction is flipped across the origin:
\begin{lstlisting}
def loss(mesh, lighting_dir):
  expr1 = infer(SoftRas(mesh, +lighting_dir))
  expr2 = infer(SoftRas(mesh, -lighting_dir))
  return -(expr1 - expr2) ** 2
\end{lstlisting}
Finally, we performed the necessary optimization by gradient descent. We ran the optimization separately from 100 random seeds. Approximately 10\% of the results were compelling to our eyes. Our top three illusions are shown in Figure~\ref{fig:face}. For these, we were able to reproduce the effect in the physical world by 3D-printing the optimized meshes and shining a flashlight on the printed objects.

Of the remaining 90\%, the most common failure mode was a mismatch between typical human face perception and our model. In unusual lighting conditions our model infers ``correct'' parameters more reliably than humans, thus missing opportunities to create illusions. Additionally, the Basel Face Model was designed using data mostly from European faces, so our generative model does not match the true distribution of faces in the world. Finally, some illusions did not work when 3D printed because of shadows, which SoftRas does not account for. We expect that by tuning the model to better match humans, one could obtain a higher illusion ``yield.''
\section{Future work}
\label{sec:future-work}

\paragraph{Discrete random variables}

Our method does not yet support discrete random variables. For example, we cannot model figure-ground reversals as in the Rubin vase illusion, because we cannot infer the binary choice between ``faces'' and ``vase.''
One promising solution is to make a differentiable version of an HMC variant like ``reflective/refractive HMC''~\cite{afshar2015reflection}, which is already used in some PPLs to allow mixing discrete and continuous random variables~\cite{zhou2019lfppl}. Alternatively, we could apply recent advances in differentiating integrals (or expectations) containing discontinuities~\cite{bangaru2021systematically}.

\paragraph{From visual illusions to visual applications}
The graphics community has long applied insights from perception to create not only visual illusions \cite{chu2010camouflage, oliva2006hybrid, chi2008self, ma2013change, yu2010embedded}, but also perceptually-inspired image processing algorithms \cite{toler2007illustration, bousseau2011optimizing, hertzmann2020line, ritschel20083d, huberman2015reducing, khan2006image, walton2021beyond}. These works span a wide variety of specialized methods, each targeting a particular aspect of perception. In this paper, we offer a single, general approach to illusion synthesis, which can be used to play with a whole host of perceptual mechanisms. While we were initially motivated by illusions, our approach should conceptually extend to any graphics task that is naturally posed as optimization over perceptual inference.

\paragraph{Beyond vision}
The ``inverse inverse'' view of depiction straightforwardly extends to Bayesian models of cognition \emph{beyond} perception, such as models of intuitive physics \cite{battaglia2013simulation} and social cognition \cite{baker2009action}. In future work, we thus envision posing animation as ``inverse inverse simulation,'' storytelling as ``inverse inverse planning,'' and so on.
More generally, we view differentiable PPLs as a new tool in the computational cognitive science toolbox, providing gradients that help probe human cognition in the same way that we currently probe neural networks.
\section{Conclusion}
\label{sec:conclusion}

We introduced a differentiable probabilistic programming language, which enables efficiently backpropagating through MCMC inference. Then, we used our new tool to generate a variety of visual illusions by ``inverse inverse graphics'': that is, by modeling vision as Bayesian inference and using gradient descent to find ``adversarial examples'' that fool those models.
\begin{acks}
We thank the anonymous reviewers for their feedback, as well as Gregory Valiant for early discussions about ``The Dress,'' Max Siegel and Thomas O'Connell for advice on conducting psychophysical experiments, and Liane Makatura for assistance with 3D printing. \linebreak
This research was supported by NSF Grants \#2105806, \#CCF-1231216, \#CCF-1723445 and \#CCF-1846502, ONR Grant \#00010803, the Hertz Foundation, the Paul \& Daisy Soros Fellowship for New Americans, and an NSF Graduate Research Fellowship under Grant \#1745302.
\end{acks}

\bibliographystyle{ACM-Reference-Format}
\bibliography{references}

\end{document}